\documentstyle[sprocl]{article}

\def\simgt{\stackrel{>}{{}_\sim}}

\def\NPB{{\em Nucl. Phys.} B}
\def\PLB{{\em Phys. Lett.}  B}

\def\PRD{{\em Phys. Rev.} D}

\def\be{\begin{equation}}
\def\ee{\end{equation}}
\def\beq{\begin{equation}}

\def\eeq{\end{equation}}
\def\bea{\begin{eqnarray}}
\def\eea{\end{eqnarray}}

\begin{document}

CERN-TH/98-41 $ $ $ $ 
OUTP-98-04


\title{RECENT DEVELOPMENTS IN ELECTROWEAK BARYOGENESIS\footnote{Talk given at the XXXIIIrd Recontres de Moriond Conference, {\it Fundamental Parameters in Cosmology}, Les Arcs, France, January 17-24, 1998.}}

\author{A. RIOTTO }

\address{CERN Theory Division,\\
CH-1211 Geneva 23, Switzerland\footnote{On leave of absence from Theoretical Physics Dept., University of Oxford, Oxford, U.K.}}


\vskip 1cm

\maketitle\abstracts{We discuss some recent developments made in the computation of the baryon asymmetry generated at the electroweak scale. We emphasize that the local number  density asymmetries of the particles involved in supersymmetric electroweak baryogenesis must be described by a  set of Quantum Boltzmann Equations.  These diffusion equations  automatically and self-consistently incorporate  the CP-violating sources which fuel baryogenesis and manifest ``memory'' effects which are typical of the quantum transport theory and are not present in the classical approach.}

\section{Prologo}

Because of  the presence of  unsuppressed
baryon number violating processes at high temperatures, 
the Standard Model (SM) of weak interactions fulfills all the requirements for  a successful   
generation of the baryon number at the electroweak scale. The baryon number violating processes    also impose severe constraints on  
models where the baryon asymmetry is created at energy scales much higher than the electroweak scale. Unfortunately, the electroweak phase transition  is too weak  in the SM. This   means that the baryon asymmetry  
generated during the transition would subsequently be  erased by unsuppressed  
sphaleron transitions in the broken phase. 
 
The most promising and  
well-motivated framework for electroweak baryogenesis beyond the SM  seems to be supersymmetry (SUSY).  Electroweak  
baryogenesis in the framework of the Minimal Supersymmetric Standard Model  
(MSSM) has  attracted much attention in the past years, with 
particular emphasis on the strength of the phase transition ~\cite{early1,early2,early3} and  
the mechanism of baryon number generation \cite{nelson,noi,higgs,ck}.

Recent  analytical \cite{r1,r2} and  lattice  
computations  \cite{r3} have also  revealed   that the phase transition can be sufficiently strongly  
first order if  the
ratio of the vacuum expectation values of the two neutral Higgses $\tan\beta$  
is smaller than $\sim 4$. Moreover, taking into account all the experimental bounds 
as well as those coming from the requirement of avoiding dangerous
 color breaking minima,  the lightest Higgs boson should be  lighter than about  $105$ GeV,
 while the right-handed stop mass might  be close to the present experimental bound and should
 be smaller than, or of the order of, the top quark mass \cite{r2}. 

Moreover, the MSSM contains additional sources  
of CP-violation  besides the CKM matrix phase. 
These new phases are essential  for the generation of the baryon number since  large  
CP-violating sources may be  locally induced by the passage of the bubble wall separating the broken from the unbroken phase during the electroweak phase transition.   Baryogenesis is fuelled  when transport properties allow the CP-violating  
charges to efficiently diffuse in front of the advancing bubble wall where
anomalous electroweak baryon violating processes are not suppressed.
The new phases   
appear    in the soft supersymmetry breaking  
parameters associated to the stop mixing angle and to  the gaugino and  
neutralino mass matrices; large values of the stop mixing angle
are, however, strongly restricted in order to preserve a
sufficiently strong first order electroweak phase transition. 
Therefore, an acceptable baryon asymmetry from the stop sector
may only be generated through a delicate balance between the values
of the different soft supersymmetry breaking parameters contributing
to the stop mixing parameter, and their associated CP-violating 
phases \cite{noi}. As a result, the contribution to the final baryon asymmetry from the stop sector turns out to be negligible.   On the other hand, charginos and neutralinos may be responsible for the observed baryon asymmetry if   the phase of the parameter $\mu$ is    large enough \cite{noi,ck}. Yet, 
this is true within the MSSM. If the strength of the
 electroweak phase transition is enhanced by the presence of some
new degrees of freedom beyond the ones contained in the MSSM, {\it
e.g.} some extra standard model gauge singlets,
 light stops (predominantly the 
right-handed ones) and charginos/neutralinos are expected to 
give quantitatively the
same contribution to the final baryon asymmetry.  

\section{The old wisdom}

In this talk we will mainly concern ourselves with the recent developments done  in the computation of the baryon asymmetry. Since it  not our intention to get into technical details, we will only try to give the feeling of what is going on.  

Let us first mention what was done in the old days.  

The baryon asymmetry has been usually computed using the following  separate steps \cite{nelson,noi}:

{\it 1)}  Look for 
 those  charges which are
approximately conserved in the symmetric phase, so that they
can efficiently diffuse in front of the bubble where baryon number
violation is fast, and non-orthogonal to baryon number,
so that the generation of a non-zero baryon charge is energetically
favoured.
Charges with these characteristics in the MSSM 
are the axial stop charge and the Higgsino charge,
which may be produced from the interactions of squarks and
charginos
and/or neutralinos with the bubble wall,
provided a source of CP-violation is present in these sectors; 

{\it 2)}  Compute the CP-violating currents  of the plasma locally induced by the passage of the bubble wall. The methods present in the literature properly incorporate the decoherence effects which may have a crucial impact on the generation of the CP-violating observable; 

{\it 3)} Write and solve a set of coupled differential diffusion equations for the local particle densities, including the CP-violating source terms derived from the computation of the current at  step {\it 2)}  and the particle number changing reactions. The solution to these equations gives a net baryon number which is produced in the symmetric phase and then transmitted into the interior of the bubbles of the broken phase, where it is not wiped out if the first transition is strong enough. 
It is important to notice that the CP-violating sources have been usually  inserted into the diffusion equations by hand only after the CP-violating currents have been defined and computed. 
This procedure is  certainly appropriate to describe the damping effects on the CP-violating observables 
originated by the
plasma interactions, but it 
does not  incorporate any relaxation time scale
arising when diffusion and particle changing interactions are
included (even though this approximation might be good if the diffusion time scales are larger than the  damping time scales) and is theoretically {\it not} consistent. More important, since a certain degree of  arbitrariness is present in the way the CP-violating sources may be defined, different  CP-violating sources have been adopted  for the stop and the Higgsino sectors in the literature \cite{nelson,noi}. This is certainly not an academic question since different sources may lead to   different numerical results for the final baryon asymmetry (especially if  the sources are  expressed in terms of   a different number of derivatives of the Higgs bubble wall profile and, therefore, in terms of  different powers of the bubble wall velocity $v_\omega$ and bubble wall width $L_\omega$). 

\section{The new wisdom}

It is indisputable  that a set of  transport (diffusion) equations {\it already} incorporating the CP-violating sources in a self-consistent way may be obtained  only by means of a more complete treatment of the problem.   Non-equilibrium Quantum Field Theory  provides us with   the necessary    tools  to  write down a set of quantum Boltzmann equations (QBE's) describing the local particle densities and automatically incorporating the CP-violating sources.    The most appropriate extension of the field theory
to deal with these issues it to generalize the time contour of
integration to a closed time-path (CTP).  The CTP formalism is a powerful Green's function
formulation for describing non-equilibrium phenomena in field theory and it leads to 
a complete
non-equilibrium quantum kinetic theory approach. 

\subsection{Out-of-equilibrium field theory with a broad rush}

The CTP formalism (often  dubbed as in-in formalism) is a powerful Green's function
formulation for describing non-equilibrium phenomena in field theory.  It
allows to describe phase-transition phenomena and to obtain a
self-consistent set of quantum Boltzmann equations.
The formalism yields various quantum averages of
operators evaluated in the in-state without specifying the out-state. On the contrary, the ordinary quantum field theory (often dubbed as in-out formalism) yields quantum averages of the operators evaluated  with an in-state at one end and an out-state at the other. 

The partition function in the in-in formalism for a {\it complex} scalar field is defined to be
\begin{eqnarray}
Z\left[ J, J^{\dagger}\right] &=& {\rm Tr}\:\left[ T\left( {\rm exp}\left[i\:\int_C\:\left(J\phi+J^{\dagger}\phi^{\dagger} \right)\right]\right)\rho\right]\nonumber\\
&=& {\rm Tr}\:\left[ T_{+}\left( {\rm exp}\left[ i\:\int\:\left(J_{+}\phi_{+}+J^{\dagger}_{+}\phi^{\dagger}_{+} \right)\right]\right)\right.
\nonumber\\
&\times&\left.  T_{-}\left( {\rm exp}\left[ -i\:\int\:\left(J_{-}\phi_{-}+J^{\dagger}_{-}\phi^{\dagger}_{-} \right)\right]\right) \rho\right],
\end{eqnarray}
where the suffic $C$ in the integral denotes that the time integration contour runs from minus infinity to plus infinity and then back to minus infinity again. The symbol $\rho$ represents the initial density matrix and the fields are in the Heisenberg picture  and  defined on this closed time contour. 

As with the Euclidean time formulation, scalar (fermionic) fields $\phi$ are
still periodic (anti-periodic) in time, but with
$\phi(t,\vec{x})=\phi(t-i\beta,\vec{x})$, $\beta=1/T$.
The temperature appears   due to boundary
condition, but time is now  explicitly present in the integration
contour.

For non-equilibrium phenomena and as a consequence of the time contour, we must now identify field
variables with arguments on the positive or negative directional
branches of the time path. This doubling of field variables leads to
six  different real-time propagators on the contour \cite{chou}.  It is possible to employ fewer  than six propagators since they are not independent, but using six simplifies the notation. 
For a generic bosonic charged  scalar field $\phi$ they are defined as 
\begin{eqnarray}
\label{def1}
G_{\phi}^{>}\left(x, y\right)&=&-i\langle
\phi(x)\phi^\dagger (y)\rangle,\nonumber\\
G_{\phi}^{<}\left(x,y\right)&=&-i\langle
\phi^\dagger (y)\phi(x)\rangle,\nonumber\\
G^t _{\phi}(x,y)&=& \theta(x,y) G_{\phi}^{>}(x,y)+\theta(y,x) G_{\phi}^{<}(x,y),\nonumber\\
G^{\bar{t}}_{\phi} (x,y)&=& \theta(y,x) G_{\phi}^{>}(x,y)+\theta(x,y) G_{\phi}^{<}(x,y), \nonumber\\
G_{\phi}^r(x,y)&=&G_{\phi}^t-G_{\phi}^{<}=G_{\phi}^{>}-G^{\bar{t}}_{\phi}, \:\:\:\: G_{\phi}^a(x,y)=G^t_{\phi}-G^{>}_{\phi}=G_{\phi}^{<}-G^{\bar{t}}_{\phi},
\end{eqnarray}
where the last two Green functions are the retarded and advanced Green functions respectively and $\theta(x,y)=\theta(t_x-t_y)$ is the step function. Analogous formulae hold for fermion fields. The reader is referred to \cite{bau} for more technical details. 

For interacting systems whether in equilibrium or not, one must define and calculate self-energy functions. There are six of them: $\Sigma^{t}$, $\Sigma^{\bar{t}}$, $\Sigma^{<}$, $\Sigma^{>}$, 
$\Sigma^r$ and $\Sigma^a$. There are some   relationships  among them, such as
\begin{equation}
\Sigma^r=\Sigma^{t}-\Sigma^{<}=\Sigma^{>}-\Sigma^{\bar{t}}, \:\:\:\:\Sigma^a=\Sigma^{t}-\Sigma^{>}=\Sigma^{<}-\Sigma^{\bar{t}}. 
\end{equation}
The self-energies are incorporated into the Green functions through the use of  Dyson's equations which are the starting point to get the Quantum Boltzmann equations. The latter look as follow \cite{bau}
\begin{eqnarray}
\label{aaa}
\frac{\partial n_\phi(X)}{\partial T}&+&\stackrel{\rightarrow}{\nabla}_{R}\cdot 
\vec{j}_\phi(X)=-\int\: d^3 r_3\:\int_{-\infty}^{T}\: dt_3\left[\Sigma^{>}_{\phi}(X,x_3) G^{<}_{\phi}(x_3,X)\right.\nonumber\\
&-&\left. G^{>}_{\phi}(X,x_3) \Sigma^{<}_{\phi}(x_3,X)
+ G^{<}_{\phi}(X,x_3)\Sigma^{>}_{\phi}(x_3,X)\right.\nonumber\\
&-&\left.\Sigma^{<}_{\phi}(X,x_3) G^{>}_{\phi}(x_3,X)\right].
\end{eqnarray}
Here $n_\phi$ and $\vec{j}_\phi$ are 
the particle density and current asymmetry, respectively,  and the right-hand side represents the ``scattering'' term. It contains  all the information necessary to describe the temporal evolution of the particle density asymmetries:  particle number
changing reactions and CP-violating source terms,  which will  pop out from the corresponding self-energy $\Sigma_{{\rm CP}}$. In supersymmetric baryogenesis
one has to consider the Quantum Boltzmann equations for right-handed stops and 
higgsinos \cite{bau}.

\subsection{It is worth doing out-of-equilibrium field theory}

One of the merits of the CTP formalism is to guide us towards a rigorous and self-consistent definition of the CP-violating sources  {\it within}  the quantum Boltzmann equations \cite{bau}.  On the contrary, previous treatments \cite{nelson,noi}  are characterized by the  following common feature:  
 CP-violating currents  were first derived and then  converted    into sources for the diffusion equations.   More specifically, CP-violating sources ${\cal S}$ associated to a generic charge density $j^0$ were constructed from 
the current $j^\mu$  by the definition ${\cal S}=\partial_0 j^0$ \cite{nelson,noi}.  A rigorous computation of the CP-violating currents for the right-handed stop $\widetilde{t}_R$ and higgsino $\widetilde{H}$  local densities  was  performed in \cite{noi} by means of the CTP formalism. Since   currents  were proportional to the first time derivative of the the Higgs profile,  sources turned out to be   proportional to the second time derivative of the Higgs profile \cite{noi}.  The results obtained in \cite{bau}, however, indicate that the sources in the quantum diffusion equations are proportional to the first time derivative of the Higgs configuration.   A comparison between the sources ${\cal S}$ obtained in \cite{bau} and the 
currents $j^0$ given in ref. \cite{noi} indicate that  they may be related
as
\begin{equation}
\label{z}
{\cal S}(T)\sim \frac{j^0(T)}{\tau}
\end{equation}
and it may be interpreted as the time derivative of the current density accumulated at the time
$T$ after  the wall has deposited at a given specific point the current density $j^0$ each interval $\tau$
\begin{equation}
{\cal S}(T)\sim \partial_0 \int^T dt \:\frac{j^0(t)}{\tau}.
\end{equation}
Here 
 $\tau=\Gamma^{-1}$ is the thermalization time of the right-handed stops and higgsinos, respectively. The integral over time is peculiar of the quantum approach and it induces memory effects. 
This tells us that the source obtained self-consistently in the present work differs from the one adopted in \cite{noi} by a  factor $\sim  L_\omega\Gamma/v_\omega$ (in the rest frame of the advancing bubble wall). Since 
$L_\omega\Gamma/v_\omega\simgt 1$  for  the Higgs derivative expansion to hold, this result is important as far as the numerical estimate of the final baryon number is concerned.   

\subsection{Memory effects}

Baryogenesis is fuelled  when transport properties 
allow the CP-violating  
charges to efficiently diffuse in front of the advancing bubble wall where
anomalous electroweak baryon violating processes are not suppressed.
However, the CP-violating processes of quantum interference,  which build up  
CP-violating sources,  must  act in opposition to  the incoherent nature of  
plasma physics  responsible for the loss of quantistic interference.  
If the particles  involved in the process of baryon number generation 
thermalize rapidly, CP-violating   
sources 
loose their coherence and are diminished. The  CTP formalism   properly describes the quantum nature of CP-violation and tells us that   
CP-violating sources evaluated at some time $T$  are always  proportional to an integral over the past history of the system.  Therefore, it is fair to  argue that these memory effects  lead to ``relaxation'' times  for the CP-violating sources which   are   typically longer than the ones dictated by the thermalization rates of the particles in the  thermal bath.  In fact, this observation
is valid for all the processes described by the ``scattering'' term in the right-handed side of the quantum diffusion equations. 
The slowdown of the relaxation processes  may help to keep the system out of equilibrium for longer times and therefore enhance the final baryon asymmetry. There are  two more  reasons why one should expect  quantum relaxation times to be  longer than the ones predicted by the classical approach. First, the decay of the Green's functions  as functions of the difference of the time arguments;    secondly,   the rather different oscillatory behaviour of the functions $G^{>}$ and $G^{<}$ present in the CTP formalism \cite{bau} for a given momentum, as functions of the time argument difference. 

\subsection{Resonance effects}

In the limit of thick bubble walls,   the CP-violating sources are characterized by resonance effects \cite{noi} when the particles involved in the construction of the source are degenerate in mass.  The interpretation of the resonance is rather straightforward if we think  in terms of  scatterings of the quasiparticles off the advancing low momentum bubble wall configuration.  A similar effect has been found in ref. \cite{ck} where the system was  studied  in the classical limit. The classical treatments should provide reasonable approximations to our formulae to those particles whose wavelength
is short compared to $v_\omega/\Gamma$. On the other hand,  formulae should not agree for small $\Gamma$ because our source is dominated by particles with long wavelength. In this regime,   the classical approximation breaks down since it  requires that  the mean free path should  be larger than  the Compton
 wavelength of the underlying particle.  This is relevant because  quasiparticles with long wavelengths  give a significant contribution to CP-violating sources.

\section*{Acknowledgments}
I  would like to thank M. Carena, J. Cline,  M. Quiros and C.E.M. Wagner for useful discussions and  Rocky  Kolb for saying during his summary  talk that  I should have talked about something else. I would also like to thank all the organizers of this  conference for the nice atmosphere they have been able to create. Last, but not least, many thanks to my ski-teacher.

\end{document}